\documentclass{article}
\usepackage{spconf,amsmath,graphicx}
\usepackage{cite}

\usepackage{afterpage}
\usepackage{array,colortbl}
\usepackage{xcolor}
\usepackage[numbers,sort&compress]{natbib}
\usepackage[utf8]{inputenc} %
\usepackage[T1]{fontenc}    %
\usepackage{hyperref}       %
\usepackage{url}            %
\usepackage{booktabs}       %
\usepackage{mathtools,amssymb}
\usepackage{amsfonts}       %
\usepackage{nicefrac}       %
\usepackage{microtype}      %
\usepackage{pgfplots,pgfplotstable}
\pgfplotsset{compat=1.14}
\usepackage{array,colortbl}
\usepackage{xcolor}
\usepackage{algorithm,algorithmicx,algpseudocode}
\usepackage[capitalise]{cleveref}
\usepackage{caption}
\usepackage{graphbox}
\usepackage{placeins}
\usepackage{wrapfig}
\usepackage{etoolbox}
\usepackage{subfloat}
\usepackage{graphicx}
\usepackage[thicklines]{cancel}

\newcommand{\bI}{\mathbf{I}}

\newcommand{\bzero}{\mathbf{0}}

\newcommand{\bx}{\mathbf{x}}
\newcommand{\by}{\mathbf{y}}
\newcommand{\bz}{\mathbf{z}}

\newcommand{\bepsilon}{{\boldsymbol{\epsilon}}}

\usepackage{amsmath}
\usepackage{caption}
\usepackage{subcaption}
\usepackage{bm}

\title{CommIN: SEMANTIC IMAGE COMMUNICATIONS AS AN INVERSE PROBLEM WITH INN-GUIDED DIFFUSION MODELS}
\name{Jiakang Chen\textsuperscript{\textdagger}, Di You\textsuperscript{\textdagger}, Deniz Gündüz, Pier Luigi Dragotti}
\address{Department of Electrical and Electronic Engineering, Imperial College London, UK}

\begin{document}
\maketitle
\renewcommand*{\thefootnote}{\fnsymbol{footnote}}
\footnotetext[2]{Equal contribution} 

\begin{abstract}
Joint source-channel coding schemes based on deep neural networks (DeepJSCC) have recently achieved remarkable performance for wireless image transmission. However, these methods usually focus only on the distortion of the reconstructed signal at the receiver side with respect to the source at the transmitter side, rather than the perceptual quality of the reconstruction which carries more semantic information. As a result, severe perceptual distortion can be introduced under extreme conditions such as low bandwidth and low signal-to-noise ratio. In this work, we propose CommIN, which views the recovery of high-quality source images from degraded reconstructions as an inverse problem. To address this, CommIN combines Invertible Neural Networks (INN) with diffusion models, aiming for superior perceptual quality. Through experiments, we show that our CommIN significantly improves the perceptual quality compared to DeepJSCC under extreme conditions and outperforms other inverse problem approaches used in DeepJSCC.
\end{abstract}
\section{Introduction}
\label{sec:intro}

Shannon's separation theorem \cite{shannon1948mathematical} forms the cornerstone of modern communication systems. In a typical point-to-point transmission, the transmitter that performs the encoding process usually consists of two steps, the source encoder that removes redundant information from the source to achieve compression, and the channel encoder that adds redundancy to the compressed information to correct for errors caused by noisy communication channels. However, the theoretical optimality of designing the source encoder and channel encoder separately can only be guaranteed if there exists an infinite code length, which is not achievable in practice \cite{kostina2013lossy}, in particular under extreme bandwidth and delay constraints. In such cases, a Joint Source-Channel Coding (JSCC) approach can potentially lead to better performance by integrating source compression and channel encoding into a single step.

\begin{figure}[!htbp]\footnotesize %
\hspace{-0.26cm}
\begin{tabular}{c@{\extracolsep{0em}}c@{\extracolsep{0em}}c@{\extracolsep{0em}}c@{\extracolsep{0em}}c@{\extracolsep{0em}}c}
       
		&\includegraphics[width=0.115\textwidth]{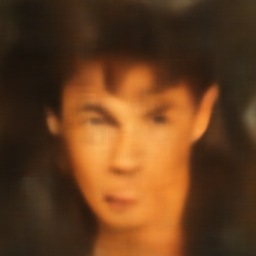}~
        &\includegraphics[width=0.115\textwidth]{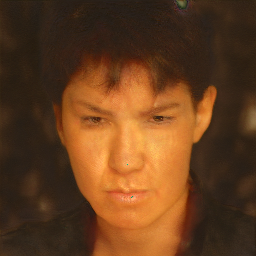}~
		&\includegraphics[width=0.115\textwidth]{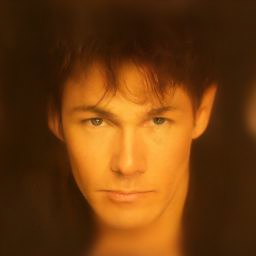}~
		&\includegraphics[width=0.115\textwidth]{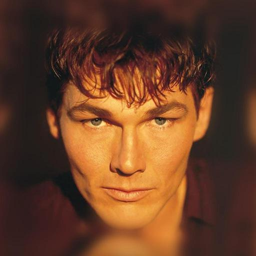}\\
        &\includegraphics[width=0.115\textwidth]{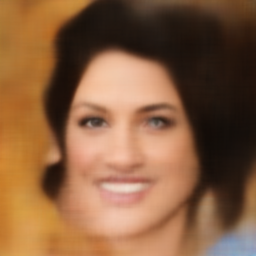}~
        &\includegraphics[width=0.115\textwidth]{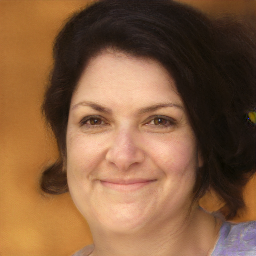}~
		&\includegraphics[width=0.115\textwidth]{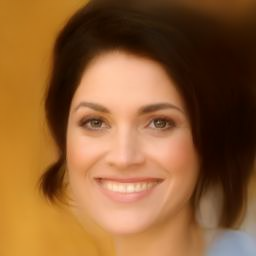}~
		&\includegraphics[width=0.115\textwidth]{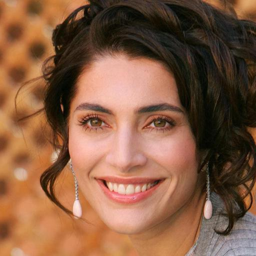}\\
   & DeepJSCC & InverseJSCC & Ours  &Ground Truth \\
	\end{tabular}
    \vspace{-0.1cm}
	\caption{Visual comparison examples of CelebA-HQ images under complex AWGN channel at SNR = -5dB and bandwidth compression ratio $\rho=0.0013$. We observe that in this very challenging communication setting, the signal reconstructed by DeepJSCC is highly distorted. While InverseJSCC improves the perceptual quality of the reconstructed images by utilizing the prior knowledge of a pre-trained StyleGAN-2 generator, our approach produces even higher perceptual quality reconstructions.}
	\label{fig:visualresults}
 \vspace{-0.5cm}
\end{figure}

\begin{figure}[t]
    \centering
    \includegraphics[width=0.49\textwidth]{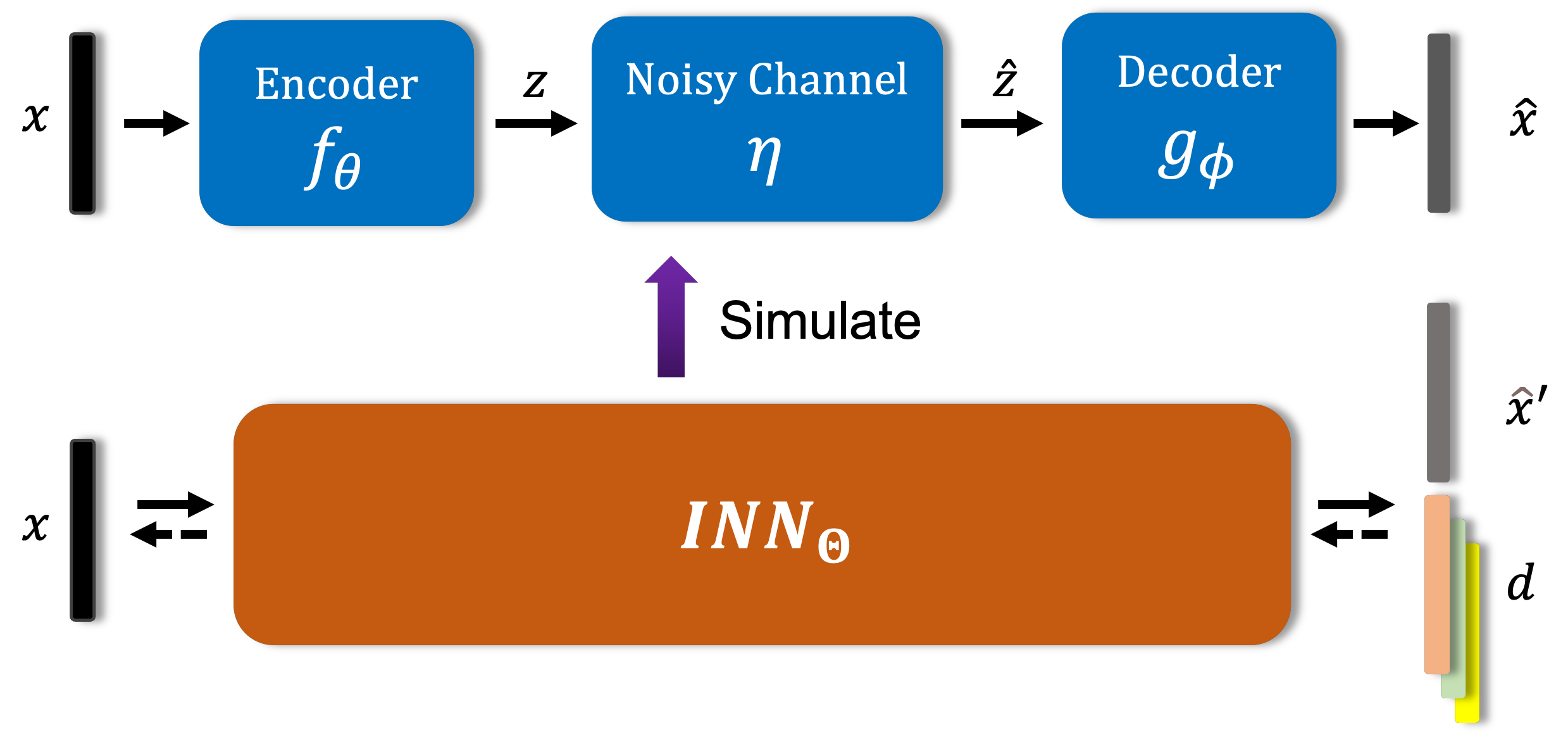}
        \vspace{-0.15cm}
\caption{ Our key insight is to simulate the degradation process of the communication pipeline (top) with INN (bottom), which can transform the input original signal $\mathbf{x}$  into the simulated degraded observation $\mathbf{\hat{x}}$ and estimate the lost details $\mathbf{d}$ by the degradation. }
\vspace{-0.3cm}
    \label{fig:idea}
\end{figure}
Recently JSCC approaches based on neural networks called DeepJSCC, have demonstrated excellent performance through powerful learning capabilities \cite{bourtsoulatze2019deep}\cite{kurka2020deep}\cite{tung2022deepjscc}\cite{tung2022deepwive}. However, traditional JSCC methods focus on pixel-level distortion or structural similarity rather than perceptual distortion, which can be measured by learned perceptual image patch similarity (LPIPS) metric \cite{lpips}. In \cite{xu2022deep}, the authors discuss the use of LPIPS directly as a loss function during training.

Generative models based on neural networks have demonstrated a strong ability to generate high perceptual quality data. Examples include variational autoencoder (VAE), generative adversarial network (GAN) and the cutting-edge diffusion models. There are several works that attempt to leverage the power of these generative models into JSCC. For example, \cite{choi2019neural} implements a JSCC scheme over binary channels via VAE. In \cite{marchioro2020adversarial}, adversarial training is employed to achieve secure JSCC in the presence of an eavesdropper. In \cite{yang2022ofdm}, GAN-style loss is employed to train the legitimate decoder. A diffusion denoising model is employed at the decoder in \cite{niu2023hybrid} as part of a hybrid transmission framework. The papers \cite{wu2023cddm} and \cite{xu2023latent} perform noise reduction by placing the diffusion model on the noisy channel symbols of JSCC to improve the reconstruction quality. However, the above approaches do not directly apply the state-of-the-art generative model structure to JSCC. Moreover, they have not fully utilised the strong prior of pretrained GANs such as StyleGAN-2, which has potential to improve the perceptual quality of the reconstruction significantly.

In \cite{bora2017compressed}, a pretrained generative model is employed to solve a classical inverse problem in compressed sensing. Outstanding performance in single-image super-resolution is achieved in \cite{menon2020pulse} through the utilization of generative models, highlighting its efficacy in addressing practical inverse problems. The similarity between communication systems and inverse problems suggests the potential of applying a powerful generative model to JSCC, as represented by works such as \cite{erdemir2022privacy}\cite{marchioro2020adversarial}\cite{genjscc}. Among them, \cite{genjscc} proposes InverseJSCC, which considers recovering high quality source images with better perceptual quality from degraded reconstructions as an inverse problem. The inverse problem is solved by a pre-trained StyleGAN-2 generator and combined with the ILO \cite{Daras2021IntermediateLO} method, which uses the state-of-the-art GAN prior to achieve a never-before-attained reconstructed perceptual quality under very extreme channel conditions.

In this paper, inspired by \cite{genjscc}, we propose CommIN, which treats the JSCC problem as an inverse problem and, by leveraging results from our recent work \cite{you2023indigo}, we reconstruct the source using Invertible Neural Networks (INN) and generative diffusion models. The key insight here is that the INN splits the signal (typically an image) into a coarse version $\mathbf{c}$ and details $\mathbf{d}$. In \cite{you2023indigo}, we train the forward part of the INN to mimic the degradation process in an inverse problem so that $\mathbf{c}$ is very close to the observed degraded image $\mathbf{y}$. In this paper we treat the communication process composed of the transmission, corruption with channel noise and reconstruction as the degradation process and mimic it using an INN (see Fig. \ref{fig:idea}). We treat the reconstructed image $\hat{\mathbf{x}}$ after transmission over a noisy channel using DeepJSCC as the degraded image in an inverse problem and apply variations of the approach in \cite{you2023indigo} using diffusion models to further improve the reconstruction.

Our overall proposed approach achieves state-of-the-art performance and has the advantage of being computationally efficient. Our primary achievements include:
CommIN stands as the first work employing INNs to simulate the degradation process in DeepJSCC.
It is also the first initiative to leverage the state-of-the-art generative prior of diffusion models to enhance the perceptual quality of DeepJSCC's reconstructions, especially in extreme conditions.

The paper is organized as follows: In Section 2, we briefly introduce the communication scenario and DeepJSCC. In Section 3, we then explain how to treat DeepJSCC as an inverse problem, and introduce the INN-Guided Diffusion model to solve this inverse problem. In Section 4, we present numerical results showing the superior performance of CommIN. Finally, Section 5 concludes the paper.

\vspace{-1pt}
\section{System Model}

We consider the point-to-point wireless transmission of images over a noisy channel (see Fig. \ref{fig:idea}). As in \cite{bourtsoulatze2019deep}, we model the transmitter and receiver as two neural networks. Specifically, the transmitter is the encoder and the receiver is the decoder of an auto-encoder system. The transmitter maps the source signal, a real vector $\mathbf{x} \in \mathbb{R}^m$ to the complex channel input vector $\mathbf{z} \in \mathbb{C}^k$ using a parameterised encoding function $\mathbf{z}=f\left(\mathbf{x} ; \boldsymbol{\theta}\right)$. Here $m$ represents the source bandwidth and $k$ the channel bandwidth. We define the bandwidth compression ratio (BCR) as $\rho = k/m$. This ratio reflects the compression applied to the signal by the communication system.
In the image transmission problem, $m=H \times W \times C$, where $H$, $W$ and $C$ are the height, width and colour channel of the input image. The channel input vector needs to satisfy the following average power constraint:
\begin{equation}
\frac{1}{k} \mathbb{E}_{\mathbf{z}}\left[\|\mathbf{z}\|_2^2\right] \leq \bar{P}.
\end{equation}
This is achieved by normalising the signal $\tilde{\mathbf{z}}$, output of the last layer of the encoder, as follows:
\begin{equation}
\mathbf{z}=\sqrt{k \bar{P}} \frac{\tilde{\mathbf{z}}}{\sqrt{{\tilde{\mathbf{z}}}^H {\tilde{\mathbf{z}}}}},
\end{equation}
where $H$ is the Hermitian transpose. Then the encoded signal $\mathbf{z}$ is sent to a noisy channel, which introduces random corruptions to the transmitted symbols. We consider a complex AWGN channel so that the received signal can be expressed as follows:
\begin{equation}
\hat{\mathbf{z}}=\eta\left(\mathbf{z}, \sigma^2\right)=\mathbf{z}+\mathbf{n}_C,
\end{equation}
where $\mathbf{n}_C$ is sampled in an independent identically distributed (i.i.d.) way from a complex Gaussian distribution with variance $\sigma^2$: $\mathbf{n}_C\sim\mathcal{C N}\left(0, \sigma^2 I_{k \times k}\right)$. The receiver uses a parameterised decoding function $\hat{\mathbf{x}}=g\left(\hat{\mathbf{z}} ; \boldsymbol{\phi}\right)$ to get the reconstruction $\hat{\mathbf{x}}$ from $\hat{\mathbf{z}}$. To represent the condition of the noisy channel, we define channel SNR as follows:
\begin{equation}
\text{Channel SNR}=10 \log _{10} \frac{\bar{P}}{\sigma^2} \mathrm{~dB}.
\end{equation}
The encoder and decoder are jointly optimised by minimising:
\begin{equation}
\arg \min _{\boldsymbol{\theta}, \boldsymbol{\phi}} \mathbb{E}_{\mathbf{x} \sim p_{\mathbf{x}}} \mathbb{E}_{\hat{\mathbf{x}} \sim p_{\hat{\mathbf{x}} \mid \mathbf{x}}}[d(\mathbf{x}, \hat{\mathbf{x}})],
\end{equation} 
where $d$ can be any distortion measure. 

 \begin{figure}[t]
    \centering
    \includegraphics[width=0.5\textwidth]{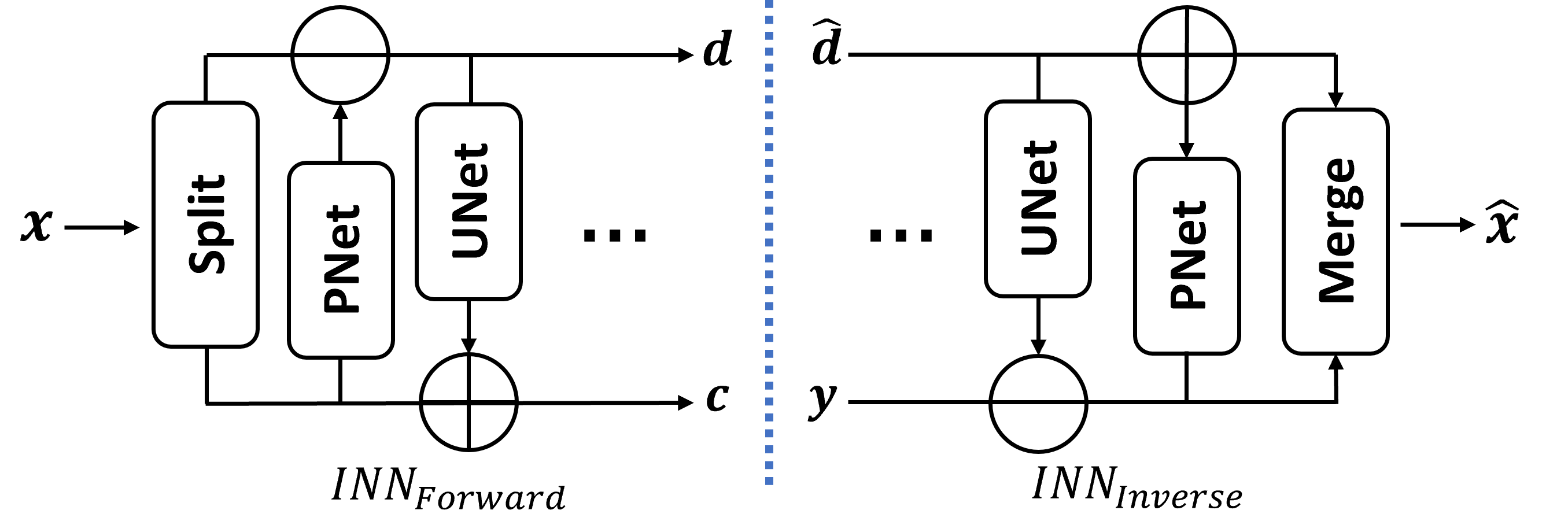}
        \vspace{-0.4cm}
\caption{Details of the framework of INN in our approach.}  \vspace{-0.25cm}
    \label{fig:Details of our INN}
\end{figure}

We use the same DeepJSCC architecture as in \cite{genjscc} but reduce the number of convolution channels from 512 to 256. The resulting DeepJSCC is optimised in an end-to-end fashion according to the Mean Squared Error (MSE) loss function:
\begin{equation}
d(\mathbf{x}, \hat{\mathbf{x}})=\frac{1}{N} \sum_{i=1}^N\left\|\mathbf{x}_i-\hat{\mathbf{x}}_i\right\|_2^2.
\end{equation}

\section{COMMIN APPROACH}
\subsection{DeepJSCC as an Inverse Problem}
Given the DeepJSCC encoder/decoder pair, we now aim to further improve the perceptual quality of the reconstructed image by treating it as the result of the degradation process in an inverse problem. Specifically, we interpret the received signal $\hat{\mathbf{x}}$ as the degraded measurements $\mathbf{y}=A(\mathbf{x})$ in a standard inverse problem, 
where $A$ is the forward operator and corresponds, in this case, to the complete communication chain of DeepJSCC including the encoder, noisy channel and decoder (see Fig.\ref{fig:idea}).
More precisely, we can express the forward operator as
\begin{equation}
\mathbf{y}=A(\mathbf{x}, \eta)
= g_\phi\left(\eta\left(f_\theta(\mathbf{x}), \sigma^2\right)\right)
=\hat{\mathbf{x}},
\end{equation}
which is a non-linear process with AWGN $\eta$.

 With the forward model in place and given measurements $\mathbf{y}$, one typically seeks to restore the original signal $\mathbf{x}$ by simultaneously imposing a fidelity constraint and a proper prior on the expected $\mathbf{x}$. The fidelity constraint is typically the MSE between the output of $A$ and the measurements $\mathbf{y}$:
\begin{equation}
\operatorname{MSE}(A(\mathbf{x}), \mathbf{y})=\|A(\mathbf{x})-\mathbf{y}\|_2^2.
\end{equation}
The prior can come in many forms and here we aim to use diffusion models as prior.

Under this setting, we are faced with two main challenges. First the degradation process is highly non-linear and cannot be expressed in closed-form. Secondly, imposing a diffusion model as prior in an inverse problem is often computationally demanding. We address these challenges by resorting to INNs. As shown in Fig.~\ref{fig:idea}, our basic idea is to simulate the degradation process by training an INN, which can decompose the original signal $\mathbf{x}$  into two parts: the estimated degraded observation $\mathbf{\hat{x}'}$ ($\approx \mathbf{\hat{x}}$) and lost details $\mathbf{d}$ during the degradation. Due to the invertibility of INN, the inverse transform of INN has the property of perfectly recovering the input original image $\mathbf{x}$ from $\mathbf{\hat{x}'}$ and $\mathbf{d}$ by construction. During the inference stage, the pre-trained INN is utilized to guide the diffusion process, ensuring that it simultaneously integrates the observed measurements $\mathbf{\hat{x}}$ and preserves the intricate details produced by the diffusion process.

In the following subsections, we first introduce the design of the INN in our algorithm, and then we describe how it works with the diffusion process to reconstruct the source signal with high perceptual quality.

\subsection{Exploring the Degradation Process with INN}

The framework of our INN is shown in Fig.~\ref{fig:Details of our INN}. The original image is split into two parts by a splitting operator. Then the Prediction Network (PNet) conditioned on the coarse part aims to predict the detail part, while the Update Network (UNet) conditioned on the detail part is used to adjust the coarse part to make it smoother. The PNet and UNet are applied alternatively to generate the coarse and detail parts, $\mathbf{c}$ and $\mathbf{d}$, respectively. The PNet and UNet can be any complex non-linear functions and their properties will not affect the invertibility of INN.

To model the degradation process,  we impose that $\mathbf{c}$ resembles $\mathbf{y}$. 
Given a training set $\left \{ \mathbf{x}^{i}, \mathbf{y}^{i}\right \}_{i=1}^{N}$, which contains $N$ high-quality images and their low-quality counterparts recovered by DeepJSCC, we optimize our INN with the following loss function: 
\vspace{-0.3cm}
\begin{align}
\begin{split}
L\left ( \Theta  \right )=\frac{1}{N}\sum_{i=1}^{N}\left \| \mathbf{c}_{i}-\mathbf{y}_{ i } \right \|_{2}^{2},
\end{split}
\end{align} 
where $\Theta$ denotes the learnable parameters of our INN.
Once we constrain one part of the output of $\textit{INN}_{Forward}(\mathbf{x})$ to be close to $\mathbf{y}$, due to invertibility, the other part of the output $\mathbf{d}$ will inevitably represent the detailed information lost during the degradation process.

\afterpage{
\begin{figure}[!tp]\footnotesize %
\hspace{-0.26cm}
\begin{tabular}{c@{\extracolsep{0em}}c@{\extracolsep{0.2em}}c@{\extracolsep{0em}}c}
       
		&\includegraphics[width=0.235\textwidth]{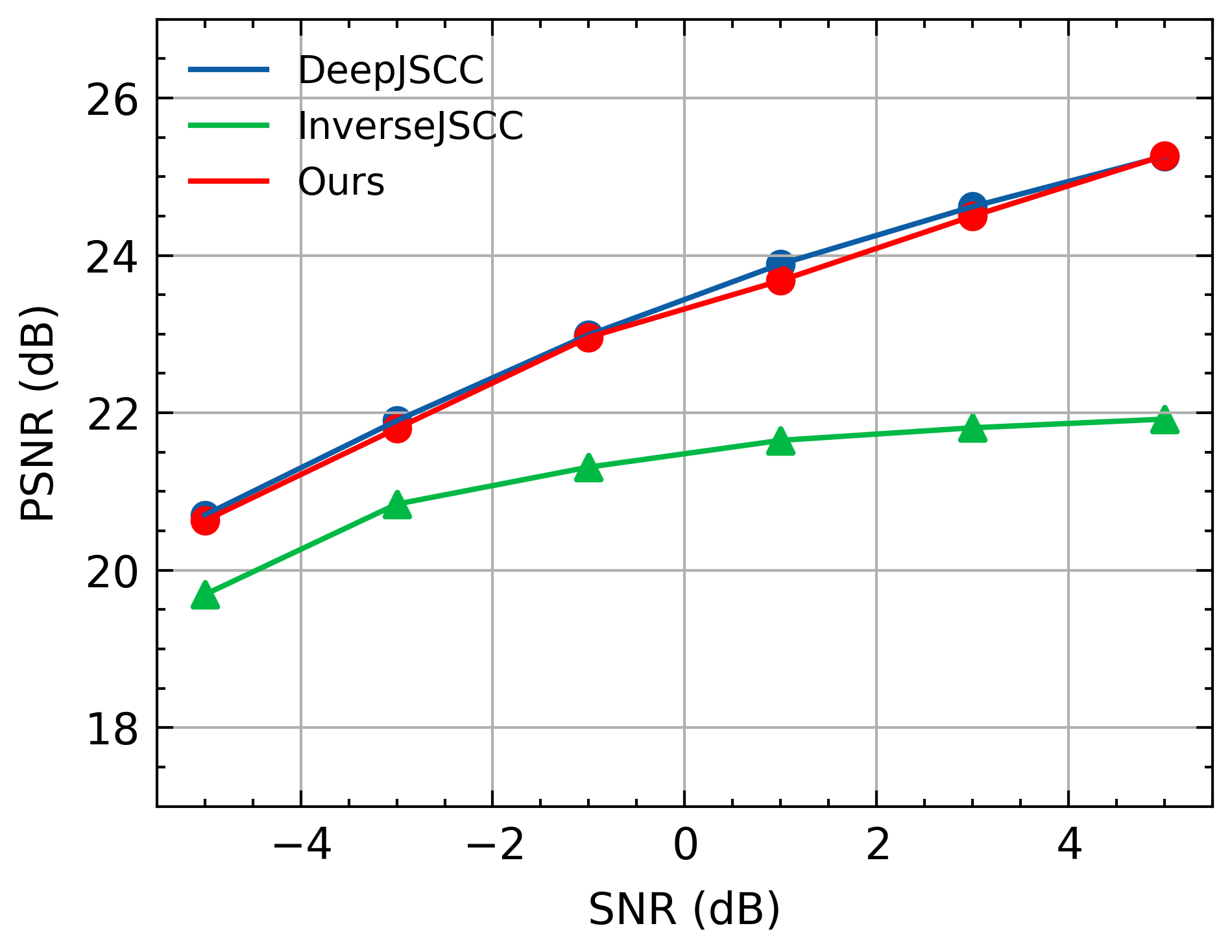}~
		&\includegraphics[width=0.235\textwidth]{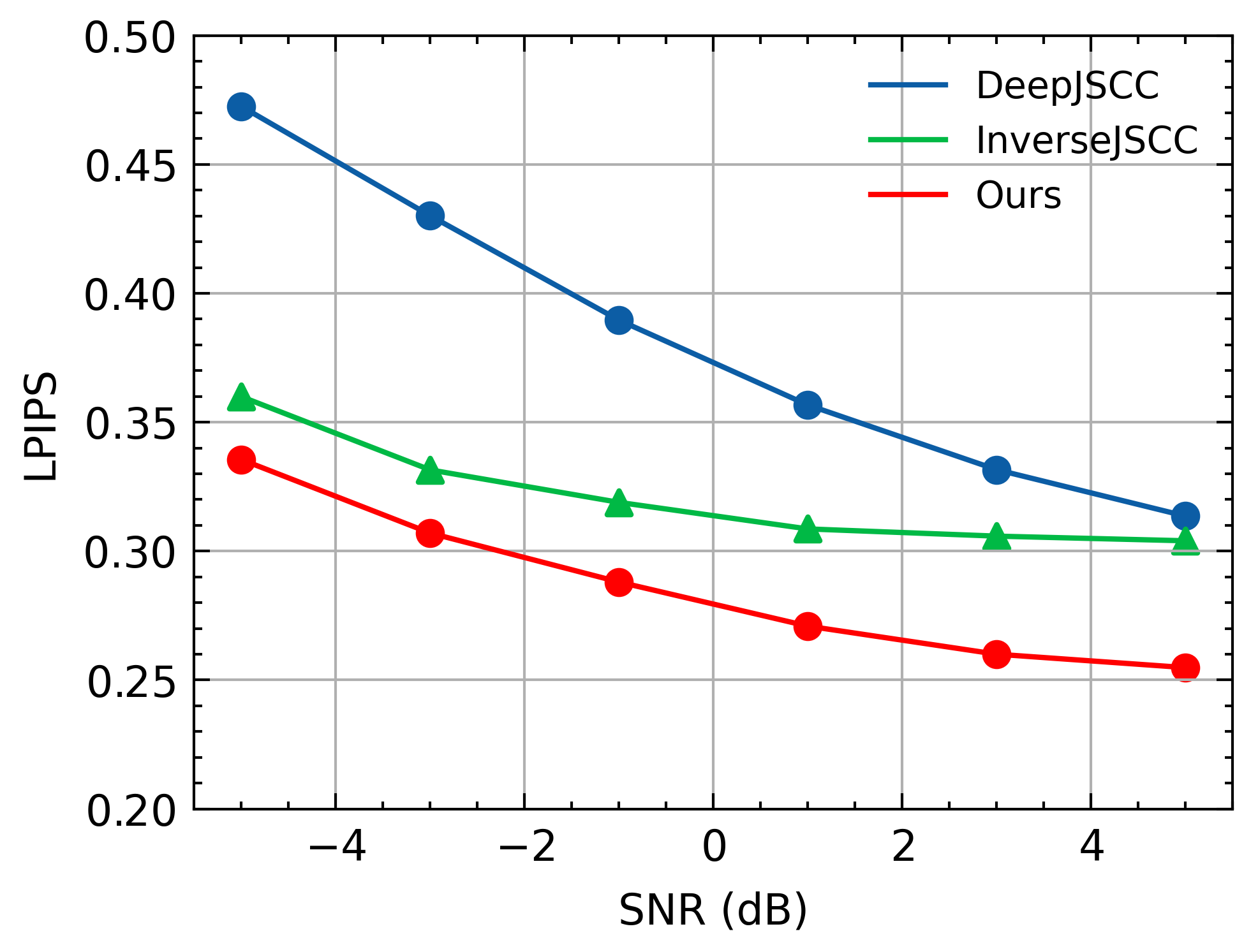}~\\
   & (a) PSNR versus SNR (higher better) & (b) LPIPS versus SNR (lower better) \\
	\end{tabular}
    \vspace{-0.1cm} 
	\caption{Performance comparison in terms of PSNR and LPIPS. } 
 \vspace{-0.1cm}
	\label{fig:performance} 
\end{figure}}
\subsection{INN-Guided Probabilistic Diffusion Algorithm}
We follow the diffusion model defined in denoising diffusion probabilistic models (DDPM \cite{ddpm}). As for other generative models, DDPM learns a distribution of images given a training set. Its inference process works by sampling a random noise vector $\mathbf{x}_{T}$ and gradually denoising it until it reaches a high-quality output image $\mathbf{x}_{0}$. Specifically, 
the reverse diffusion process iteratively samples $\mathbf{x}_{t-1}$ from $p(\mathbf{x}_{t-1}|\mathbf{x}_{t})$ to yield clean images $\mathbf{x}_{0} \sim q(\mathbf{x})$ from initial random noise $\mathbf{x}_{T} \sim \mathcal{N}(\mathbf{0},\mathbf{I})$. 
The following equations illustrate this process:
\vspace{-0.1cm}
\vspace{-0.6cm}
\begin{equation}
\mathbf{x}_{0,t}  =
\frac{1}{\sqrt{\bar\alpha_t}}(\mathbf{x}_{t} - \sqrt{1 - \bar\alpha_t} \bepsilon_\theta(\mathbf{x}_t, t) ),
\vspace{-0.5cm}
\end{equation}
and
\begin{equation}
{\color{black}\mathbf{x}_{t-1}=
{\frac{\sqrt{\alpha_t}(1-\bar\alpha_{t-1})}{1 - \bar\alpha_t}\mathbf{x}_{t}+\frac{\sqrt{\bar\alpha_{t-1}}\beta_t}{1 - \bar\alpha_t}\mathbf{x}_{0,t}  + \sigma_t \bz}},
\end{equation}
where $\bepsilon_\theta(\mathbf{x}_t, t)$ is the pre-trained denoising model and $\mathbf{x}_{0,t}$ is the estimation of  $\mathbf{x}_{0}$ from the noisy image $\mathbf{x}_{t}$. 

To solve inverse problems with the above generative model, we need to refine each unconditional transition using $\by$ to ensure data consistency. In our proposed algorithm, we impose our data-consistency step by optimizing the clean image $\mathbf{x}_{0,t}$ instead of the noisy image $\mathbf{x}_{t}$.
As shown in Algorithm~\ref{1}, we impose an additional data consistency step (in blue) after each original unconditional sampling update. At this additional step, we decompose the intermediate result $\mathbf{x}_{0,t}$ with $\textit{INN}_{Forward}$ into coarse $\mathbf{c}_{t}$ and detail part $\mathbf{d}_{t}$ and replace the coarse part $\mathbf{c}_{t}$ with the output of the DeepJSCC decoder, $\by$. The INN-optimized $\hat\bx_{0,t}$ is generated by the inverse process $\textit{INN}_{Inverse}$. Thus, the INN-optimized $\hat\bx_{0,t}$ is composed of the coarse information $\by$ and the details generated by the diffusion process. To incorporate the INN-optimized $\hat\bx_{0,t}$ into the DDPM algorithm, 
we update $\bx_{t}$ with the guidance of the gradient of $\|{\hat\bx_{0,t} - \bx_{0,t}}\|_2^2$. With the help of INN, our algorithm effectively estimates the details lost in the degradation of the communication process and avoids being limited by the requirement of knowing the closed-form expression of the degradation model.

\begin{minipage}[t]{0.44\textwidth}
    \vspace{-0.45cm}
\begin{algorithm}[H]
  \caption{DDPM Sampling with INN} \label{alg:sampling}
  \small
  \centering
  \begin{algorithmic}[1]
    \vspace{.04in}
    \State $\bx_T \sim \mathcal{N}(\bzero, \bI)$
    \For{$t=T, \dotsc, 1$}
      \State $\bz \sim \mathcal{N}(\bzero, \bI)$ if $t > 1$, else $\bz = \bzero$
      \State{{${\color{black}{\bx_{0,t}  = \frac{1}{\sqrt{\bar\alpha_t}}(\bx_{t} - \sqrt{1 - \bar\alpha_t} \bepsilon_\theta(\bx_t, t) )}}$}}
        \State{$\tilde{\bx}_{t-1}  = \frac{\sqrt{\alpha_t}(1-\bar\alpha_{t-1})}{1 - \bar\alpha_t}\bx_{t}+\frac{\sqrt{\bar\alpha_{t-1}}\beta_t}{1 - \bar\alpha_t}\bx_{0,t}  + \sigma_t \bz$}
              \State {\color{blue}$\mathbf{c}_{t},\mathbf{d}_{t}= \textit{INN}_{Forward}(\bx_{0,t})$}
            \State {\color{blue}$\hat\bx_{0,t}=\textit{INN}_{Inverse}(\by,\mathbf{d}_{t})$}  
                \State{\color{blue}{$\bx_{t-1} =\tilde{\bx}_{t-1}  - { {\zeta}}\nabla_{\bx_{t}} \|{\hat\bx_{0,t} - \bx_{0,t}}\|_2^2$}}
    \EndFor
    \State \textbf{return} $\bx_0$
    \vspace{.04in}
  \end{algorithmic}
  \label{1}
\end{algorithm}
\end{minipage}
\vspace{0.32cm}

\vspace{-0.1cm}
\section{Experiment Results}
\vspace{-0.15cm}
\label{sec:typestyle}   
To evaluate the performance of our proposed approach, we compare CommIN with state-of-the-art JSCC methods: DeepJSCC \cite{tung2022deepwive} and InverseJSCC \cite{genjscc}. We use CelebA-HQ train datasets for training. For fair comparisons, we use the same pre-trained encoder and decoder for all three approaches. 
All our experiments are implemented in PyTorch with Adam optimiser. 
In training our DeepJSCC, our channel SNR is uniformly sampled from $\left[-5, 5\right]$ dB.
We train separate INNs for each degradation setting in the inference stage.
We assume perfect channel knowledge at the encoder and the decoder. For validation, we choose CelebA-HQ 1K test datasets with image sizes 256×256, and we experiment on channel SNRs of $\{-5,-3,-1,1,3,5\}$ dB and BCR $\rho= 0.0013$, which correspond to highly challenging communication scenarios. 
Empirically, we set $T=1000$ and set step size $\zeta$ to 0.3, 0.4, and 0.5
for SNR = $\{-5, -3\}$ dB, $\{-1, 1\}$ dB, and $\{3, 5\}$ dB respectively.
We analyse the performance of the proposed scheme using both the widely used pixel-wise metric PSNR and the perceptual metric LPIPS \cite{lpips}.

\begin{figure}[!tp]\footnotesize %
\hspace{-0.26cm}
\begin{tabular}{c@{\extracolsep{0em}}c@{\extracolsep{0em}}c@{\extracolsep{0em}}c@{\extracolsep{0em}}c@{\extracolsep{0em}}c}
       
		&\includegraphics[width=0.115\textwidth]{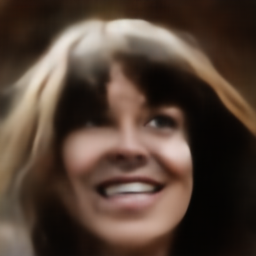}~
        &\includegraphics[width=0.115\textwidth]{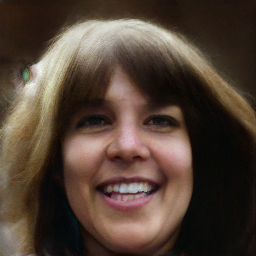}~
		&\includegraphics[width=0.115\textwidth]{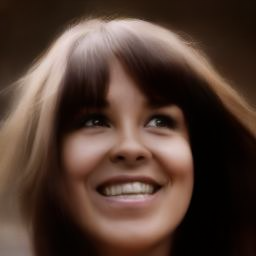}~
		&\includegraphics[width=0.115\textwidth]{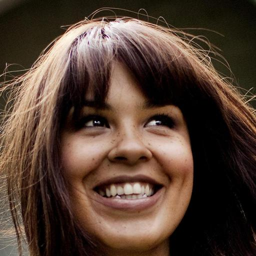}\\
        &\includegraphics[width=0.115\textwidth]{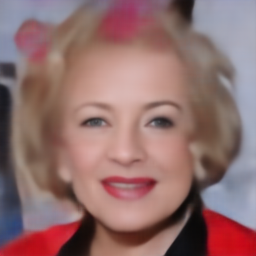}~
        &\includegraphics[width=0.115\textwidth]{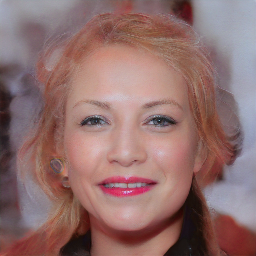}~
		&\includegraphics[width=0.115\textwidth]{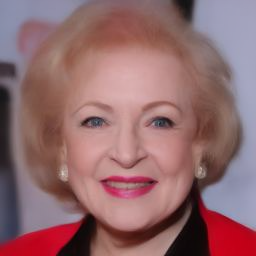}~
		&\includegraphics[width=0.115\textwidth]{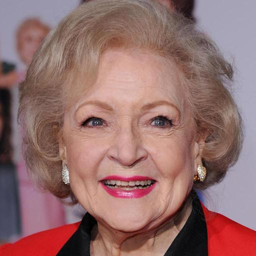}\\
   & DeepJSCC & InverseJSCC & Ours  &Ground Truth \\
	\end{tabular}
	\caption{Reconstructed CelebA-HQ images by DeepJSCC, InverseJSCC and our approach for $\rho$=0.0013 and SNR=1dB.} \vspace{-0.3cm}
	\label{fig:visual_result}
\end{figure}

Fig. \ref{fig:performance} shows our results in terms of PSNR and LPIPS with respect to channel SNR on the CelebA-HQ dataset. 
For all the settings, our approach achieves the lowest LPIPS, while maintaining the similar PSNR with DeepJSCC.
As shown in Fig. \ref{fig:visual_result}, our algorithm produces high-quality reconstruction results and preserves more details than other methods.

\vspace{-0.3cm}
\section{Conclusion}
\vspace{-0.15cm}
In this paper, we have introduced CommIN, a new JSCC scheme for wireless image transmission which treats the reconstruction of received corrupted images as an inverse problem. Key features of the approach include the use of INN to model the degradation introduced by the channel and DeepJSCC. Moreover, INNs facilitate the integration of diffusion models in the reconstruction process. 
CommIN achieves image reconstructions with perceptual quality beyond DeepJSCC in extreme environments with extemely low BCR with SNR. It also achieves competitive results compared with the existing inverse problem approach to JSCC in \cite{genjscc}.

\newpage
\bibliographystyle{ieeetr}
\bibliography{main}

\end{document}